\def\dol{{D_{\rm ol}}}
\def\dls{{D_{\rm ls}}}
\def\dos{{D_{\rm os}}}
\def\kpc{{\rm kpc}}
\def\max{{\rm max}}
\def\lsim{{ {}^<_\sim}}
\def\gsim{{ {}^>_\sim}}

\documentstyle[11pt,conf_iap]{article}
\begin{document}
\heading{THE HOLLYWOOD STRATEGY FOR 
\\ MICROLENSING DETECTION OF PLANETS } 


\author{A. GOULD$^{1}$}
       {$^{1}$ Astronomy Dept., Ohio State Univ., 174 West 18th Ave., Columbus,
OH. }

\bigskip

\begin{abstract}{\baselineskip 0.4cm 
Follow the big stars!  I review the theory of detection and parameter 
measurement of planetary systems by follow-up observations of ongoing 
microlensing events.  Two parameters can generically be measured from the 
event itself: the planet/star mass ratio, $q$, and the planet/star separation 
in units of the Einstein ring.  I emphasize the advantages of monitoring
events with giant-star sources which are brighter (thus easier to monitor) and 
bigger (thus offering the prospect of measuring an additional parameter from 
finite-source effects: the proper motion $\mu$).  There is potentially a 
strong degeneracy between $q$ and $\mu$.  I present a simple analytic 
representation of this degeneracy.  I then describe how it can be broken using 
accurate single-band photometry from observatories around the world, or
optical/infrared photometry from a single site, or preferably both.  
Both types of observations are underway or will be soon.  Monitoring of
giant-star events seen toward the bulge is also the best way to determine
the content and structure of the inner Galaxy.
}
\end{abstract}

\section{Introduction}

	Two world-wide networks are currently searching for
extra-solar planetary systems by making densely sampled observations of
ongoing microlensing events toward the Galactic bulge (PLANET \cite{PLANET}; 
GMAN \cite{GMAN}).  
Several other groups will join the search shortly and
there is serious discussion of new initiatives that would intensify the
search by an order of magnitude.  More than 100 microlensing events have
been detected to date by three groups, MACHO \cite{MACHO},
OGLE \cite{OGLE}, and DUO \cite{DUO}, based on observations made
once or twice per night.  The events typically last one week to a few months.
MACHO and OGLE have reported ``alerts'', events detected before peak.  This
alert capability 
is what has allowed PLANET and GMAN to make intensive, sometimes 
round-the-clock, follow-up observations in hopes of finding the planetary
perturbations which are expected to last a day or less.
EROS \cite{EROS} will shortly initiate a search for bulge 
microlensing events using a new 1 square-degree camera which should more
than double the number of alerts.

	In sharp contrast to this explosion of observational activity,
theoretical work on planet detection has been rather sparse, amounting to only
five papers in as many years.  Mao \&
Paczy\'nski (1991) originally suggested that planets might be detected
in microlensing events \cite{MP}.  
Gould \& Loeb (1992) developed a formalism for
understanding the character of planetary perturbations and made systematic
estimates of the rate of detection for various planetary-system parameters
\cite{GL}.
Bolatto \& Falco (1994) studied the detection rate in the more general
context of binary systems \cite{BF}.  

	Early work assumed that the lensed star could
be treated as a point source.  The usefulness of this approximation depends
primarily on $\rho$, the ratio of the angular size of the source, $\theta_*$, 
to the
planetary Einstein ring, $\theta_p$,
\begin{equation}
\rho = \frac{\theta_*}{\theta_p},\qquad
\theta_p = \sqrt{q}\theta_e,\qquad q = \frac{m}{M},\qquad
\theta_e = \sqrt{\frac{4 G M \dls}{c^2\dol\dos}}\label{eq:thetap}.
\end{equation}
Here $\theta_e$ is the Einstein ring of the lensing star, $m$ and $M$ are the
masses of the planet and its parent star, and
$\dol$, $\dls$, and $\dos$ are the distances between the observer, lens, and
source.  For Jupiter-mass planets at typical distances $(\dls\sim 2\,\kpc)$
from bulge giant sources, $\theta_p\sim 3\theta_*$, so the approximation is
a reasonable one.  However, for Saturns, Neptunes, and especially Earths,
the finite size of the source becomes quite important, and even for Jupiters
it is not completely negligible.  Moreover, as I will stress below, it is
quite possible to mistake a ``Jupiter event'' in which the source size is
negligible for a ``Neptune event'' with $\theta_*>\theta_p$.  Hence it
is essential to understand finite-source effects even to interpret events
where the source size is in fact small.

	Progress on finite-source effects was substantially delayed by 
problems of computation.  Like all binary lenses, planetary-systems have
caustics, curves in the source plane where a point-source is infinitely 
magnified as two images 
either appear or disappear.  If one attempts to integrate the magnification
of a finite source that crosses a caustic, one is plagued with numerical
instabilities near the caustic.  While it is straight forward to solve these
problems for any given geometry, the broad range of possible geometries makes
it difficult to develop an algorithm sufficiently robust for a statistical
study of lensing events.  Bennett \& Rhie (1996) solved this problem by
integrating in the image plane (where the variation of the magnification is
smooth) rather than the source plane (where it is discontinuous)
\cite{BR}.  They were
thereby able to investigate for the first time the detectability of Earth
to Neptune mass planets.  Gould \& Gaucherel (1996) showed that this approach
could be simplified from a two-dimensional integral over the image of the
source to a one-dimensional integral over its boundary \cite{GGauch}.
The difficult computational problems originally posed by 
finite-source effects are now completely solved.

\section{The Chang-Refsdal Formalism}

	Gould \& Loeb  analyzed the problem of planet detection by
treating the planet as a perturbation on a uniform background shear, $\gamma$,
produced by the star
\cite{GL}.  Let $x$ be the source-lens separation in units of 
$\theta_e$.  Then the two images are at $y_\pm$ with respectively shears
$\gamma_\pm$ and magnifications $A_\pm$, 
\begin{equation}
\gamma_\pm = y_\pm^{-2},\qquad y_\pm = \frac{\sqrt{x^2+4}\pm x}{2},
\qquad A_\pm = \frac{1}{|1-\gamma_\pm^2|}
\label{eq:gammaeq}.
\end{equation}
Since the two unperturbed images are separated by $>2\theta_e$, while the
effective range of the planet is only $\sim \theta_p$, the planet
can affect at most one image.  The Gould-Loeb approach is to treat the
other image as unperturbed with magnification given by 
equation~\ref{eq:gammaeq}
and to focus the analysis on the perturbed image.  This image is treated
as a Chang-Refsdal lens \cite{CR} \cite{SEF},
a point mass superposed on a constant background shear.  The shear
is chosen as the shear at the position of the unperturbed image at the
midpoint of the perturbation.  The actual shear due to the lensing star is,
of course, not the same at the unperturbed and perturbed positions of the
images and also varies with time during the planetary perturbation.  
Nevertheless, because the range of the planet's effect is small, the errors
induced by this approximation are negligibly small \cite{GGone}.

	The Chang-Refsdal approximation immeasurably simplifies the analysis
of planetary lensing events.  To lowest order, one measures 6 parameters
of a planetary-system light curve,
\begin{equation}
t_0,\qquad \beta,\qquad t_e,\qquad x_d,\qquad \delta_d,\qquad t_d,
\label{eq:evparms}
\end{equation}
The first three are the parameters of the unperturbed event, its time of
maximum, impact parameter in units of $\theta_e$, and Einstein crossing time.
The next three describe the gross features of the planetary perturbation,
the source-lens separation at the midpoint of the perturbation, the maximum
fractional deviation from a standard light curve, and the full-width
half-maximum (FWHM).  The value
of $\gamma$ can then be computed up to a two-fold ambiguity 
$\gamma=\gamma_\pm(x_d)$ using equation~\ref{eq:gammaeq}.  The two cases are
easily distinguished provided there is reasonable coverage of the light curve 
because perturbations of the minor $(-)$ image have a large negative excursion
surrounded by positive excursions while perturbations of the major $(+)$ 
image are positive in the middle
\cite{GL} \cite{BR} \cite{GGone}.  I focus here mainly on the major-image 
perturbation because it occurs more frequently and is somewhat easier to
understand.

	The two-dimensional structure of the magnification contours 
(up to an overall scale factor $q^{1/2}$) is fixed by this determination of 
$\gamma$ \cite{GL} \cite{GGone}.
The angle at which the source traverses this structure is given by
$\sin\phi=\beta/x_d$.  There remain three unknowns: $q$, $\rho$, and $\alpha$,
where the last is the planet/unperturbed-image separation in units of 
$\theta_p$.

\section{Point Sources}

	Suppose that the source were somehow known to be small compared
to the planet Einstein ring, $\rho\ll 1$.  If the source passed outside the
caustic region, the perturbation would be characterized by a smooth bump.
By comparing the observed height of the bump $\delta_d$ with the height
of the perturbation ridge, one could determine the planet/unperturbed-image 
separation
up to a two-fold ambiguity $\pm\alpha$.  The combination of observables
\begin{equation}
Q = \biggl(\sin\phi \frac{t_d}{2 t_e}\biggr)^2
\label{eq:qdef},
\end{equation}
could then be compared with the FWHM of the perturbation contours
at $\alpha$ to determine $q$.  For ease of illustration, I will assume here 
that this FWHM is $2\theta_p$ (which is a good general approximation).  In
any event, it is no trouble to calculate the exact value for any particular
case.  For this choice,
\begin{equation}
q = Q,\qquad \rho\ll 1
\label{eq:qrho}.
\end{equation}
The planet/star separation in units of the Einstein ring
is then $y_p = y_+(x_d)\pm \alpha q^{1/2}$.
Thus, the fractional uncertainty in the separation is $\sim 2\alpha q^{1/2}$.
For example, for $q=10^{-3}$ (as with Jupiter and the Sun) and $\alpha=10$
(which is not at all atypical) this uncertainty would be $\sim 60\%$.  This
degeneracy can be broken from the asymmetry of the perturbation.  If, for
example, the perturbation takes place after the peak of the event, then
the planet is closer to (farther from) the lens than $y_+$ if the leading
wing of the perturbation is higher (lower) than the trailing wing.

	If the source passed over or very close to the caustic, it would
be possible to determine $q$, $\rho$, and $\alpha$ just as it is for all
other binary-lens caustic-crossing events.

\section{Degeneracy From Finite Source Effects}

	If $\delta_d\ll 1$, then one possible cause is that $\alpha\gg 1$.
However, another possible cause is that $\rho\gg 1$.  In fact, for
$\alpha=0$ and a source which is larger than the caustic structure, 
Gould \& Gaucherel find analytically that the fractional deviation of the major
image is \cite{GGauch}
\begin{equation}
\delta_d = \frac{2}{\rho^2 A(\gamma)} + {\cal O}(\rho^{-4}),
\qquad A(\gamma)= \frac{1+\gamma^2}{1-\gamma^2} 
\label{eq:deltad}.
\end{equation}
[For the minor image, $\delta_d\sim {\cal O}(\rho^{-4})$.]  If $\rho\gsim1$, 
then
the duration of the perturbation is set by the size of the source, not the
planet Einstein ring.  The net result is that the solution
\begin{equation}
q\sim  \frac{Q}{\rho_\max^2},\qquad \rho\sim
\rho_\max= \sqrt{\frac{2}{\delta_d}\,
\frac{1-\gamma^2}{1+\gamma^2}}
\label{eq:qrhoalt},
\end{equation}
reproduces the peak and FWHM just as well as equation~\ref{eq:qrho}.  
Intermediate solutions are also allowed.  Since one hopes to detect planetary
systems with deviations at least as small as $\sim 5\%$, this degeneracy
can be rather severe.  For example, for $\gamma=0.6$, the range of allowed
masses is $\sim \delta_d^{-1}$ or $\sim 20$ for $\delta_d\sim 5\%$.  For
high mass planets ($\rho<1$) some events will have $\delta_d\sim 1$ in
which case there is no mass degeneracy.  However, most will have $\delta_d<1$
and, if the degeneracy is not broken, these can be confused with planets
of much lower mass ratios.  Moreover, for low mass planets, one always has 
$\delta_d<1$, so these can never be unambiguously identified without
breaking the degeneracy.

\section{Breaking the Mass Degeneracy}

	There are essentially two methods for breaking this degeneracy:
detailed light curves and optical/infrared photometry.  If a point source
$(\rho\ll 1)$ passes far from the planet $(\alpha\gg 1)$, then the wings of
the deviation will show a smooth rise and fall.  On the other hand, if a
large source ($\rho\gsim 1$) passes over the caustic $(\alpha\lsim 1)$, 
the leading
wing will show a slight fall and then an abrupt rise as the source passes
over regions of negative deviation and then infinite magnification.  The
trailing wing shows the same behavior in reverse.  The difference between
these two curves is typically of order a few per cent and is concentrated
in two brief intervals (typically a few hours) on either side of the peak.
Thus, both accurate photometry and good weather at the appropriate observing 
station at the right time are required.

	A second method is to use simultaneous optical/infrared (e.g. $V$ and
$H$) photometry.  Since giant stars are more limb-darkened at bluer 
wavelengths, the light curve will show color variations if the source passes
over (and is therefore resolved by) a caustic.  When the red leading edge
passes over the caustic, the image becomes red.  At the peak it becomes
blue, then red again.  For a point source passing far from the planet, there
are no color changes.  Since the color changes are smaller, $\sim 1-2\%$, this
method requires even better photometry.  Moreover, it requires specialized
equipment in order to observe simultaneously in optical and infrared light.
However, the color variations occur at all phases of the deviation, so no
special observing luck is required.  Finally, the specialized camera has
been approved by the US NSF and should be ready by the bulge season of 1998
(private communication, D.\ DePoy 1996).  Probably, the best solution is
to combine continuous round-the-clock coverage with optical/infrared
photometry from at least one site.

	If the degeneracy is broken, then one measures at least two parameters,
$q$ and $y_p$.  If the source passes over the caustic, or if $\rho>1$, then
one measures a third parameter, $\rho$.  Note that in this case, one also
determines the proper motion $\mu$,
\begin{equation}
\mu = \frac{\theta_*}{\rho t_e\sqrt{q}}
\label{eq:mueq}.
\end{equation}

\section{From Mass Ratios to Masses}

	From $t_e$ alone, one can estimate $M$ and $r_e=\dol\theta_e$ 
only to about
a factor of 3.  Consequently, even if $q$ and $y_p$ are determined 
unambiguously, the mass $m=q M$ and physical projected separation
$a_p=r_e y_p$ are in general only known to a factor 3.  If
$\mu$ is measured, these uncertainties are reduced to about a factor 2.
As mentioned above, $\mu$ can be measured for most planetary events for
which $\rho\gsim 1$.  
Using equation~\ref{eq:thetap}, one finds that this occurs
provided
\begin{equation}
m > 30\,M_\oplus\,\frac{\dol/\dls}{3}\,\biggl(\frac{r_*}{10\,r_\odot}\biggr)^2
\label{eq:rhoeq},
\end{equation}
where $r_*$ is the radius of the source.
Recall that it can also be measured for larger planets provided that the 
source crosses the caustic.

	If a parallax satellite \cite{Refsdal} \cite{Gone} were launched, it
could routinely measure the projected Einstein radius $\tilde r_e =
(\dos/\dls)r_e$ \cite{GGtwo}.  This by itself would dramatically
reduce the uncertainties in $M$ and $r_e$ \cite{HG} and so in
$m$ and $a_p$.  Moreover, if $\mu$ (or $\rho$) were also measured, the planet 
mass and projected separation would be determined \cite{Gtwo},
\begin{equation}
m = \frac{c^2}{4 G}\frac{\theta_*\tilde r_e\sqrt{q}}{\rho},\qquad
a_p=\frac{y_p}{\tilde r_e^{-1}+ \rho\sqrt{q}/r_*}
\label{eq:degenbrk}.
\end{equation}

\section{Giants Rule}

	For some time, I have been advocating that microlensing searches
toward the bulge focus primarily on giants \cite{Gtwo}.  This view is
motivated primarily by a desire to understand the structure and content
of the Galaxy.  Giant events can be detected in $R$ band even for
$A_R=4.5$ ($A_V=5.7$), and so could be found in all but the most heavily
extincted regions of the bulge.  They are not crowded, so they are hardly
affected by blending.  In addition, they suffer no significant bias toward
monitoring stars on the near side of the bulge compared to the far side.
These factors mean that the observed optical depths and time
scales of detected events do not require large and uncertain corrections
as a function of position.  In addition, the blending suffered by turnoff stars
generates a large tail of spurious short events \cite{Han} and even if one
had confidence in the statistical corrections to this effect, it would be
difficult to unambiguously detect or rule out a large brown dwarf population
in the bulge using these sources.  Finally, it is possible to measure the
proper motions of a much bigger fraction of giant star than turnoff events
both because they are larger (and so more often give rise to finite source
effects) and they are brighter, especially in the infrared where their
two images can sometimes be resolved using interferometry \cite{Gthree}.

	Are giant sources also better for planet searches?  The question is
important because the same follow-up observations are used both to detect
planets and probe Galactic structure.  I believe the answer to this question
is:  Yes, definitely.  First, giants are brighter and hence easier to
monitor.  Since dozens of stars must be monitored simultaneously, bright
sources (and hence short exposures) are highly valued.  Second, giants can
be seen even in the relatively heavily obscured central portions of the 
Galaxy where there are likely to be more lensing events.  Third, while it is
true that the peak deviation is suppressed when a giant is lensed by a small
planet, the onset of this effect is only at $m\lsim 30\,M_\oplus$ for 
planets in
the bulge, and $m\lsim 10\,M_\oplus$ for planets in the disk 
(see eq.~\ref{eq:rhoeq}).  Moreover, suppression for these small-mass planets
does not become severe until one reaches masses that are a factor
3--10 lower.  And these events with suppressed peaks have
several compensating advantages 
including measurement of $\mu$ and a higher event rate due to a larger
cross section.  It is true that for Earth-mass planets, the suppression
can be so severe that the event is missed altogether \cite{BR}.  
Thus, to detect Earth-mass planets, it may in fact be necessary to monitor
turn-off stars.  The cross sections for Earth-mass planets are so low that
there are probably not enough giants available to monitor to detect 
Earth-mass planets anyway.  Hence, they
require an order-of-magnitude larger search than is likely to be
conducted in the near future, with more and larger follow-up telescopes so
that many hundreds of faint-star lensing events can be followed on one-hour
time scales.  For the present, giant sources are the indicated choice.
\bigskip

\acknowledgements{I would like to thank Scott Gaudi for many useful 
discussions and Dimitar Sasselov for bringing to my attention the Hollywood
connection.  This work was supported by grant AST 94-20746 from the NSF.}

\newpage

%

%
\vfill

\begin{thebibliography}{99}{\baselineskip 0.4cm
\bibitem{DUO} Alard, C.\ 1996. in Proc. IAU Symp.\ 173 
{\it Astrophysical Applications of Gravitational Lensing}
p.\ 215, eds.\ C.\ S.\ Kochanek, J.\ N.\ Hewitt), Kluwer Academic Publishers 
\bibitem{PLANET} Albrow, M., et al.\ 1996. in Proc. IAU Symp.\ 173 
{\it Astrophysical Applications of Gravitational Lensing}
p.\ 227, eds.\ C.\ S.\ Kochanek, J.\ N.\ Hewitt), Kluwer Academic Publishers 
\bibitem{MACHO} Alcock, C.\  1995. \apj\, {\bf 445}, 133
\bibitem{EROS} Aubourg, E., et al.\ 1995. \aa\, {\bf 301}, 1
\bibitem{BR} Bennett, D.\ P., \& Rhie, S.\ H.\ 1996. \apj\, {\it in press}
\bibitem{BF} Bolatto, A.\ D., \& Falco, E.\ E.\ 1994. \apj\, {\bf 436}, 112
\bibitem{CR} Chang, K., \& Refsdal, S.\ 1979. \nat\, {\bf 282}, 561
\bibitem{GGone} Gaudi, B.\ S., \& Gould, A.\ 1996. \apj\, {\it in preparation}
\bibitem{Gone} Gould, A.\ 1995. \apj\, {\bf 441}, L21
\bibitem{Gtwo} Gould, A.\ 1995. \apj\, {\bf 447}, 491
\bibitem{Gthree} Gould, A.\ 1996. {\it PASP}\, {\bf 108}, 465
\bibitem{GGauch} Gould, A., \& Gaucherel, C.\ 1996. \apj\, {\it submitted}
\bibitem{GGtwo} Gould, A., \& Gaudi, B.\ S.\ 1997. \apj\, {\it almost in press}
\bibitem{GL} Gould, A., \& Loeb, A.\ 1992. \apj\, {\bf 396}, 104
\bibitem{Han} Han, C.\ 1996. \apj\, {\it submitted}
\bibitem{HG} Han, C.\ \& Gould, A.\ 1995. \apj\, {\bf 447}, 53
\bibitem{MP} Mao, S., \& Paczy\'nski, B.\ 1991. \apj\, {\bf 374}, L37
\bibitem{GMAN} Pratt, M., et al.\ 1996. in Proc. IAU Symp.\ 173 
{\it Astrophysical Applications of Gravitational Lensing}
p.\ 221, eds.\ C.\ S.\ Kochanek, J.\ N.\ Hewitt), Kluwer Academic Publishers 
\bibitem{Refsdal} Refsdal, S.\ 1966. \mn\, {\bf 134}, 315
\bibitem{SEF} Schneider, P., Ehlers, J., \& Falco, E.\ E.\ 1992. {\it
Gravitational Lenses}, Springer-Verlag
\bibitem{OGLE} Udalski, A., et al.\ 1994. {\it Acta Astronomica}\, {\bf 44}, 
165
}
\end{thebibliography}
\end{document}